\documentclass{article}
\pdfoutput=1
\usepackage{amssymb}
\usepackage{amscd}
\usepackage{amsmath}
\usepackage{graphicx}

\begin{document}

%%%%%%%%%%%%%%%%%%%%%%%%%%%%%%%%%%%%

\newcounter{mo}
\newcommand{\mo}[1]
{\stepcounter{mo}$^{\bf MO\themo}$%
\footnotetext{\hspace{-3.7mm}$^{\blacksquare\!\blacksquare}$
{\bf MO\themo:~}#1}}

\newcounter{bk}
\newcommand{\bk}[1]
{\stepcounter{bk}$^{\bf BK\thebk}$%
\footnotetext{\hspace{-3.7mm}$^{\blacksquare\!\blacksquare}$
{\bf BK\thebk:~}#1}}

%%%%%%%%%%%%%%%%%%%%%%%%%%%%%%%%%%%%

\newcommand{\Si}{\Sigma}
\newcommand{\tr}{{\rm tr}}
\newcommand{\ad}{{\rm ad}}
\newcommand{\Ad}{{\rm Ad}}
\newcommand{\ti}[1]{\tilde{#1}}
\newcommand{\om}{\omega}
\newcommand{\Om}{\Omega}
\newcommand{\de}{\delta}
\newcommand{\al}{\alpha}
\newcommand{\te}{\theta}
\newcommand{\vth}{\vartheta}
\newcommand{\be}{\beta}
\newcommand{\la}{\lambda}
\newcommand{\La}{\Lambda}
\newcommand{\D}{\Delta}
\newcommand{\ve}{\varepsilon}
\newcommand{\ep}{\epsilon}
\newcommand{\vf}{\varphi}
\newcommand{\vfh}{\varphi^\hbar}
\newcommand{\vfe}{\varphi^\eta}
\newcommand{\fh}{\phi^\hbar}
\newcommand{\fe}{\phi^\eta}
\newcommand{\G}{\Gamma}
\newcommand{\ka}{\kappa}
\newcommand{\ip}{\hat{\upsilon}}
\newcommand{\Ip}{\hat{\Upsilon}}
\newcommand{\ga}{\gamma}
\newcommand{\ze}{\zeta}
\newcommand{\si}{\sigma}

\def\hS{{\hat{S}}}

\newcommand{\li}{\lim_{n\rightarrow \infty}}
\def\mapright#1{\smash{
\mathop{\longrightarrow}\limits^{#1}}}

\newcommand{\mat}[4]{\left(\begin{array}{cc}{#1}&{#2}\\{#3}&{#4}
\end{array}\right)}
\newcommand{\thmat}[9]{\left(
\begin{array}{ccc}{#1}&{#2}&{#3}\\{#4}&{#5}&{#6}\\
{#7}&{#8}&{#9}
\end{array}\right)}
\newcommand{\beq}[1]{\begin{equation}\label{#1}}
\newcommand{\eq}{\end{equation}}
\newcommand{\beqn}[1]{\begin{small} \begin{eqnarray}\label{#1}}
\newcommand{\eqn}{\end{eqnarray} \end{small}}
\newcommand{\p}{\partial}
\def\sq2{\sqrt{2}}
\newcommand{\di}{{\rm diag}}
\newcommand{\oh}{\frac{1}{2}}
\newcommand{\su}{{\bf su_2}}
\newcommand{\uo}{{\bf u_1}}
\newcommand{\SL}{{\rm SL}(2,{\mathbb C})}
\newcommand{\GLN}{{\rm GL}(N,{\mathbb C})}
\def\sln{{\rm sl}(N, {\mathbb C})}
\def\sl2{{\rm sl}(2, {\mathbb C})}
\def\SLN{{\rm SL}(N, {\mathbb C})}
\def\SLT{{\rm SL}(2, {\mathbb C})}
\def\PSLN{{\rm PSL}(N, {\mathbb C})}
\newcommand{\PGLN}{{\rm PGL}(N,{\mathbb C})}
\newcommand{\gln}{{\rm gl}(N, {\mathbb C})}
\newcommand{\PSL}{{\rm PSL}( 2,{\mathbb Z})}
\newcommand{\SLZ}{{\rm SL}(2, {\mathbb Z})}
\def\f1#1{\frac{1}{#1}}
\def\lb{\lfloor}
\def\rb{\rfloor}
\def\sn{{\rm sn}}
\def\cn{{\rm cn}}
\def\dn{{\rm dn}}
\newcommand{\rar}{\rightarrow}
\newcommand{\upar}{\uparrow}
\newcommand{\sm}{\setminus}
\newcommand{\ms}{\mapsto}
\newcommand{\bp}{\bar{\partial}}
\newcommand{\bz}{\bar{z}}
\newcommand{\bw}{\bar{w}}
\newcommand{\bA}{\bar{A}}
\newcommand{\bG}{\bar{G}}
\newcommand{\bL}{\bar{L}}
\newcommand{\btau}{\bar{\tau}}

\newcommand{\tie}{\tilde{e}}
\newcommand{\tial}{\tilde{\alpha}}

\newcommand{\Sh}{\hat{S}}
\newcommand{\vtb}{\theta_{2}}
\newcommand{\vtc}{\theta_{3}}
\newcommand{\vtd}{\theta_{4}}

\def\mC{{\mathbb C}}
\def\mZ{{\mathbb Z}}
\def\mR{{\mathbb R}}
\def\mN{{\mathbb N}}

\def\frak{\mathfrak}
\def\gb{{\frak b}}
\def\gg{{\frak g}}
\def\gp{{\frak p}}
\def\gn{{\frak n}}
\def\gJ{{\frak J}}
\def\gS{{\frak S}}
\def\gL{{\frak L}}
\def\gM{{\frak M}}
\def\gG{{\frak G}}
\def\gE{{\frak E}}
\def\gF{{\frak F}}
\def\gk{{\frak k}}
\def\gK{{\frak K}}
\def\gl{{\frak l}}
\def\gh{{\frak h}}
\def\gH{{\frak H}}
\def\gs{{\frak s}}
\def\gt{{\frak t}}
\def\gu{{\frak u}}
\def\gT{{\frak T}}
\def\gR{{\frak R}}

\def\baal{\bar{\al}}
\def\babe{\bar{\be}}

\def\bfa{{\bf a}}
\def\bfb{{\bf b}}
\def\bfc{{\bf c}}
\def\bfd{{\bf d}}
\def\bfe{{\bf e}}
\def\bff{{\bf f}}
\def\bfg{{\bf g}}
\def\bfm{{\bf m}}
\def\bfn{{\bf n}}
\def\bfp{{\bf p}}
\def\bfu{{\bf u}}
\def\bfv{{\bf v}}
\def\bfr{{\bf r}}
\def\bfs{{\bf s}}
\def\bft{{\bf t}}
\def\bfx{{\bf x}}
\def\bfy{{\bf y}}
\def\bfM{{\bf M}}
\def\bfR{{\bf R}}
\def\bfC{{\bf C}}
\def\bfP{{\bf P}}
\def\bfq{{\bf q}}
\def\bfS{{\bf S}}
\def\bfJ{{\bf J}}
\def\bfI{{\bf I}}
\def\bfX{{\bf X}}
\def\bfY{{\bf Y}}
\def\bfz{{\bf z}}
\def\bfnu{{\bf \nu}}
\def\bfsi{{\bf \sigma}}
\def\bfU{{\bf U}}
\def\bfso{{\bf so}}

\def\clA{\mathcal{A}}
\def\clC{\mathcal{C}}
\def\clD{\mathcal{D}}
\def\clE{\mathcal{E}}
\def\clF{\mathcal{F}}
\def\clG{\mathcal{G}}
\def\clR{\mathcal{R}}
\def\clU{\mathcal{U}}
\def\clT{\mathcal{T}}
\def\clO{\mathcal{O}}
\def\clH{\mathcal{H}}
\def\clK{\mathcal{K}}
\def\clJ{\mathcal{J}}
\def\clI{\mathcal{I}}
\def\clL{\mathcal{L}}
\def\clM{\mathcal{M}}
\def\clN{\mathcal{N}}
\def\clP{\mathcal{P}}
\def\clQ{\mathcal{Q}}
\def\clV{\mathcal{V}}
\def\clW{\mathcal{W}}
\def\clZ{\mathcal{Z}}

\def\baf{{\bf f_4}}
\def\bae{{\bf e_6}}
\def\ble{{\bf e_7}}
\def\bag2{{\bf g_2}}
\def\bas8{{\bf so(8)}}
\def\baso{{\bf so(n)}}

\def\sr2{\sqrt{2}}
\newcommand{\ran}{\rangle}
\newcommand{\lan}{\langle}
\def\f1#1{\frac{1}{#1}}
\def\lb{\lfloor}
\def\rb{\rfloor}
\newcommand{\slim}[2]{\sum\limits_{#1}^{#2}}

%\def\theequation{\thesubsection.\arabic{equation}}% the equation
               % number now does not include the section number;
               % \setcounter{equation}{0} should be put after every
               % \section{} command!!!
\newcommand{\sect}[1]{\setcounter{equation}{0}\section{#1}}
\renewcommand{\theequation}{\thesection.\arabic{equation}}
\newtheorem{predl}{Proposition}[section]
\newtheorem{defi}{Definition}[section]
\newtheorem{rem}{Remark}[section]
\newtheorem{cor}{Corollary}[section]
\newtheorem{lem}{Lemma}[section]
\newtheorem{theor}{Theorem}[section]

\vspace{0.3in}

 \begin{flushright}
 ITEP-TH-26/23\\
 IITP-TH-20/23
 \end{flushright}
\vspace{1.cm}
%\today

\vspace{10mm}

%\begin{document}

%%%%%%%%%%%%%%%%%%%%%%%%%%%%%%%%%%%%%%%%%%%%%%%%%%%%%%%%%%%%%%%%%%%%%%%%%%%%%%%%%%%%%
\begin{center}
{\LARGE{\bf Two types of Witten zeta functions }}\\
\vspace{12mm}
%\large {
 A. Levin$^{a\,b}$
% {\Large {
  M.Olshanetsky$^{b\,c}$
   \\
 \vspace{1mm}
\vspace{1mm}$^a$ - 
%{\sf
 Laboratory of Algebraic Geometry, GU-HSE
 \\
\vspace{1mm}$^b$ - 
%{\sf
National Research Centre "Kurchatov Institute"
% NRC "Kurchatov Institute", 123182, Moscow, Russia
 \\
\vspace{1mm}$^c$ -  Institute for Information Transmission Problems, Moscow 127994, Russia

\end{center}
\vspace{4mm}
\today
\vspace{4mm}
\begin{abstract}
We define two types of Witten's zeta functions
according to Cartan's classification of compact symmetric spaces.
The type II is the original Witten zeta function constructed by means
 of irreducible representations of the simple compact Lie group U.
 The type I Witten zeta functions, we introduce here, are related to
  the irreducible spherical representations of U. They arise in the
  harmonic analysis on  compact symmetric spaces of the form U/K, where
   K is the maximal subgroup of U.
  To construct the type I zeta function we calculate
   the partition functions of 2d YM theory with broken gauge symmetry
    using the Migdal-Witten approach. We prove that for the rank one
symmetric spaces the generating series for
the values of  the type I functions with integer arguments can be defined in terms of
the generating series of the Riemann zeta-function.

%We define two types of.
%We introduce the type I Witten. While the Witten zeta functions are related
%to the set of irreducible representations of simple compact Lie groups,
%the type I Witten zeta functions are related to the irreducible spherical representations
%of  compact Lie groups.
%They are related to the compact symmetric spaces of the form $U/K$, where U is a compact
%simple group and $K$ its maximal subgroup.
%%form a special subclass  of irreducible representations.
%%of simple compact Lie groups.
%  Simple compact groups are  a special class of compact symmetric spaces.
%According to the Cartan classification compact groups are
% they are
% called the type II symmetric spaces. Therefore, we call the Witten zeta functions
% are the functions of type II. The coset spaces  are called the type I symmetric spaces.

%Another class of compact symmetric spaces are the coset spaces $X=U/K$,
% where $K$ is a maximal compact subgroup of $U$. These
%The  Witten zeta function of type I is the main object of our investigation.
\end{abstract}

 \footnotetext{E-mails:
{\em alevin57@gmail.com}; {\em olshanet@itep.ru}}

\section{Introduction}

  Let $U$ be a compact simple Lie group and $\Si_g$ is a curve of genus $g$.
 Witten in \cite{Wi} calculated the volume of the moduli space of flat $U$-bundles over
  $\Si_g$. To do this he constructed the partition function of  2d Yang-Mills theory.
  The calculations in the 2d Yang-Mills theory had been done previously \cite{F,Mi,Ru}.
  A crucial point in Witten's calculations is the  A. Migdal's lattice approximation
  of the functional integral. As a result
  the partition function takes the form of series over dominant weights
  $\G^+(U)$ of irreducible representations of $U$.
 As the coupling constant of  YM theory tends to zero this series becomes what Zagier
  called the Witten zeta function (WZF) \cite{Za}.
  It can be regarded as a group theoretical generalisation of the
   Riemann zeta function,  since
   for $U=SU(2)$ the WZF coincides with it.
 %Riemann zeta function $\zeta(s)$, for $U=$SO$(3)$ with the Hurwitz zeta function $\zeta(s,\oh)$,
 Note that for $U=$SU$(3)$ it is related to the  Mordell-Tornheim zeta function.
 In recent years, the Witten zeta function has been the subject of muny mathematical works
 \cite{Ma}.

  Simple compact groups are  a special class of compact symmetric spaces.
According to the Cartan classification of symmetric spaces \cite{He1} they are
 called the type II symmetric spaces. Therefore, we call these zeta functions type II.
Another class of compact symmetric spaces are the coset spaces $X=U/K$,
 where $K$ is a maximal compact subgroup of $U$. These are called the type I symmetric spaces.
The  Witten zeta function of type I is the main object of our investigation.
It is defined by series over the dominant class-one representations $\G^+(U/K)$ associated with
 the pair $(U,K)$.
With this construction we
 bring the type I WZF into correspondence with some symmetric spaces  such as
  spheres, projective spaces, Grassmannians and so on (see Tables 1 and 2 below). In particular, the type I
WZF of the sphere $S^2$ is the  Hurwitz zeta function $\zeta(s,\oh)$.
To construct the type I WZF we calculate partition functions of 2d YM theory with broken gauge symmetry using the Migdal-Witten approach. Finally, we prove that for the rank one
symmetric spaces the type I the generating series for
the values of WZF with integer arguments can be defined in terms of
the generating series of the Riemann zeta-function.
%%The close to this result was  obtained in \cite{Car}, where the authors proved that

 It should be noted that there are  closed objects to the Witten zeta functions.
 The Witten zeta functions are related to the degeneracy of the spectrum of the
Laplace-Beltrami operators on a compact symmetric spaces.
Another zeta functions related to the spectrum of these operators are the Minakshisundaram-Pleijel zeta-functions \cite{MP}.
The properties of these functions have been studied in \cite{Ca,Cam,Ik,T}.

The paper is organized as follows. In  next section we explain briefly Migdal-Witten  construction. In Section 3 we define the type I
models. In Section 4 we consider the class one representations. They determine the type I zeta functions.
  Finally,  in Section 5 we investigate in details the rank one symmetric spaces and construct
  the generating functions for the values of the type I zeta functions at integer arguments.

%%%%%%%%%%%%%%%%%%%%%%%%%%%%%%%%%%%%%%%%%%%%%%%%%%%%%%%%%%%%%%%%%%%%%%%%%%%%%%%%%%%%%%%%%%%%%
%%%%%%%%%%%%%%%%%%%%%%%%%%%%%%%%%%%%%%%%%%%%%%%%%%%%%%%%%%%%%%%%%%%%%%%%%%%%%%%%%%%%%%%%%%%%%%%%%%%

\section{2d Yang-Mills theory and Witten zeta functions}

In this section we briefly recall Witten's construction of zeta functions associated with simple compact groups.
Let $U$ be a compact simple Lie group and  $\clP$ is a a principal $U$-bundle over $\Si_g$.
Consider the 2d Yang-Mills $U$ theory on the  curve $\Si_g$ \cite{F,Mi,Ru,Wi}.
It means that we
 consider an adjoint-bundle $\ad(\clP)$ over $\Si_g$ equipped with connection
$d+A$.
The 2d YM  action has the form
\beq{act}
I(e)=-\f1{2e^2}\int_{}d\mu\,\tr (F,F)\,,
\eq
where $d\mu$ is a measure on $\Si_g$ and $F$ is the curvature divided on the volume two-form
on  $\Si_g$.

The partition function is defined by the functional integral
\beq{fi}
Z( U,\Si_g,e)=\int DA\exp-I(e)\,.
\eq
This functional integral was defined
 by a lattice approximation  by Migdal \cite{Mi} for $\mR^2$  and
by Witten \cite{Wi} for general Riemann surface. It takes the following form.
Let  $R=\{\al\}$ be a system of roots corresponding to the group $U$ and $\G(U)=\{\nu\}$ is the weight lattice of irreducible unitary representations of $U$ in the finite-dimensional spaces $V_\nu$
\beq{gu}
\G(U)=\{\nu\in\gh^*\,|\,2(\nu,\al)/(\al,\al)\in\mZ\}
\eq
 and  $\G^+(U)$ is the subset of the dominate weights
 \beq{dgu}
 \G^+(U)=\{\nu\in\gh^*\,|\,2(\nu,\al)/(\al,\al)\in\mZ^+\}\,.
 \eq
 The partition function of theory takes the form
\beq{1a}
Z^{II}( U,\Si_g,q)=\sum_{\nu\in \G^+(U)}\frac{q^{c_2(\nu)}}{(d^{II}_\nu)^{2g-2}}\
\eq
Here $c_2(\nu)$ is the eigenvalue of the second Casimir operator, $q=\exp(-\varrho)$, $\varrho=Vol\,(\Si_g)$, and $d_\nu=\dim\,V_\nu$
$$
 d_\nu=\prod_{\al\in R^+}d_\al(\nu)\,,~~d_\al(\nu)=
\frac{(\nu+\rho,\al)}{(\rho,\al)}
$$
For $q=1$ $(\varrho\to 0)$
\beq{z2}
Z^{II}(U,\Si_g,1)=\sum_{\nu\in \G^+(U)}\frac{1}{(d_\nu)^{2g-2}}
\eq
 is the volume of the moduli space $\clM(U,\Si_g)$ of flat $U$-bundles
over $\Si_g$ \cite{Wi}.
The generalization of (\ref{z2})
\beq{wi}
\zeta^{II}_U(s)=\sum_{\nu\in \G^+(U)}\frac{1}{(d_\nu)^s}\,,~~s\in\mC\,.
\eq
is called by Zagier the\emph{ Witten zeta-functions} \cite{Za}. In particular,
 $\zeta^{II}_{SU_2}(s)$ is the Riemann zeta-function. We introduce  the superscript II to distinguish it from the defined below type I zeta-functions (\ref{wi1}).

%%%%%%%%%%%%%%%%%%%%%%%%%%%%%%%%%%%%%%%%%%%%%%%%%%%%%%%%%%%%%%%%%%%%%%%%%%%%
\subsubsection{Brief explanation of (\ref{1a})}
%The replacement of $Z( U,\Si_g,e)$ (\ref{fi}) to  (\ref{1a}) was described by Migdal
% for $\mR^2$ and in
%general case by Witten \cite{Wi}. The integral (\ref{fi})  can be defined
%rigorously by a lattice approximation of $\Si_g$.
To define (\ref{1a}), we  cover $\Si_g$ with polygons. In the limit as the
lattice becomes finer, the desired continuum theory can be recovered.

Let $\ga$  enumerates the  edges of polygons and $i$ run over all plaquettes (the interior part of the polygon). Fix a plaquette $w$ and let $u_{\ga_i}=u_i$ be a map of the edge $\ga_i$ of
the plaquette $w$ to the group $U$.
%The gauge group is the map of the set of vertices $V_a$
%of the lattice to the group $U$. Let $(a,b)$ be two ends of the edge $\ga$.

 Consider the holonomy
 $$
 \clU_w=u_1\ldots u_k
  $$
   around the plaquette $w$. The lattice gauge transformations acts on   $\clU_w$
by the conjugation. In particular,
the trace $\tr(\clU_w)$ is gauge invariant. It is a function on the group $U$
and
%invariant under the conjugation. Therefore
 it can be expanded in the basis of
characters $\chi_\nu(u)$ of irreducible representations.  According with \cite{Mi,Wi}
define the function on the plaquette $w$
\beq{mp}
\mathcal{G}^{II}_w(\clU_w,\varrho_w)=\mathcal{G}^{II}(u_i,\varrho_i)=\sum_{\nu\in  \G^+(U)}d_\nu\chi_\nu(\clU_w)\exp(-\varrho_w c_2(\nu)/2)\,,
\eq

The functional integral (\ref{fi}) is approximated by the sum
\beq{fa}
Z^{II}(U,\Si_g)=
\int \prod_\ga du_\ga\prod_i\mathcal{G}^{II}(u_i,\varrho_i)\,.
\eq

This formula can be justified by two arguments.
First, consider its behavior for a small plaquette as $\varrho_w\to 0$.
Since the delta functional
$\int_Udu\de(u)f(u)=f(Id)$ has the expansion
\beq{dca}
\de(u)=\sum_{\nu\in \G^+(U)}d_\nu\chi_\nu(u)
\eq
the factor $\mathcal{G}_w(\clU_w,\varrho_w)$ provides the correct continuum limit
$$
\lim_{\varrho_w\to 0}\mathcal{G}^{II}_w(\clU_w,\varrho_w)=\de(\clU_w-Id)\,.
$$

%The main justification for (\ref{mp}) is that
Moreover, the factor $\mathcal{G}_w(\clU_w,\varrho_w)$  defines a theory invariant
under subdivision. The proof of this fact in \cite{Wi}  is based on the orthogonality relation for the characters
\beq{o1}
\int_U\chi_\nu(u)\overline{\chi}_\mu(uv^{-1})du=\de_{\mu\nu}\f1{d_\nu}\chi_\nu(v)\,.
\eq
We also need another important  properties of characters (see section 3.3.1)
\beq{f5}
 \int_U\chi_\nu(uAu^{-1}B)du=\f1{d_\nu}\chi_\nu(A)\chi_\nu(B)\,.
\eq
\beq{f51}
 \int_U\chi_\mu(Au)\chi_\nu(u^{-1}B)du=\de_{\mu\nu}\f1{d_\nu}\chi_\nu(AB)\,.
\eq

%Summation over plaquettes gives the partition function (\ref{1}).
From (\ref{mp}) we obtain
\beq{fa1}
Z^{II}( U,\Si_g)=\sum_{\nu\in  \G^+(U)}d_\nu e^{-\rho c_2(\nu)/2}
\int du_idv_j\chi_\nu(u_1v_1u_1^{-1}v_1^{-1}\cdots u_gv_gu_g^{-1}v_g^{-1})\,.
\eq
It follows from (\ref{f5}) that every time one integrates over $u_i$ or $v_j$ one gets a factor of $1/d_\nu$.
On the last step we come to the integral
$$
%\f1{d^{2g-1}_nu}
\int_Udu_g\chi_\nu(u_g)\chi_\nu(u_g^{-1})=\chi_\nu(1)=d_\nu\,.
$$
Putting it all together we come to (\ref{1a}).

%%%%%%%%%%%%%%%%%%%%%%%%%%%%%%%%%%%%%%%%%%%%%%%%%%%%%%%%%%%%%%%%%%%%%%%%%%%%%%%%%%

\section{Type I models}

Let $K$ be a maximal subgroup of $U$.
The representation of $U$ in the space $V$ is called \emph{the class one or spherical} with respect
to $K$ if $V$  contains a $K$ invariant vector \cite{He2,Vi,Wa}.
 We denote by $\G(U/K)$ the sublattice of the class one weights $\G(U/K)\subset\G(U)$ and by
 $\G^+(U/K)$ the subset of the dominant weights (see below (\ref{co})).

 %%%%%%%%%%%%%%%%%%%%%%%%%%%%%%%%%%%%%%%%%%%%%%%%%%%%%%%%%%%%%%%%%%%%%%%%%%
 \subsubsection{Type I zeta functions}

The  Witten zeta function of type I is defined as
\beq{wi1}
\zeta^I_{U/K}(s)=\sum_{\nu\in  \G^+(U/K)}\frac{1}{(d_\nu)^s}\,,~~d_\nu=\dim\,V_\nu\,,
\eq
where the dimension $d_\nu$ is defined below (\ref{dim1}). However, there is another expressions
for $d_\nu$, $\,\nu\in \G^+(U/K)$ (\ref{di4}), (\ref{di5}), (\ref{nm}), and  (\ref{3d}), which we will use in this paper.
It follows from the general formulae (\ref{di}), (\ref{mr}) that $d^I_\nu$ is a polynomial
$P_N(\nu)$
of rank$(U/K)$-variables $(\nu_1,\ldots,\nu_{rank})$, degree $N=(\dim(U/K)-rank(U/K))$
and $P_N(0)=1$.
The difference with the type II Witten zeta function is that summation goes over
class one dominate weights and in the forms of $d^I_\nu$ and $d_\nu^{II}$.

To define the zeta-function (\ref{wi1}) we consider below the partition function $Z^I$.
 Then, as above the lattice approximation of
$Z^I$
leads to the expressions (\ref{fa3}), (\ref{z1}). After its calculation we obtain
\beq{5}
Z^I( U,\Si_g,q)=\sum_{\nu\in \G^+(U/K)}\frac{q^{c_2(\nu)}}{(d_\nu)^{2g-1}}\,.
\eq
 In the limit $q\to 1$ ($\varrho\to 0$) we come to the type I zeta function (\ref{wi1}).
\beq{lim}
\zeta^I_{U/K}(s)|_{s=2g-1}=\lim_{q\to 1} Z^I( U,\Si_g,q)\,.
\eq

%%%%%%%%%%%%%%%%%%%%%%%%%%%%%%%%%%%%%%%%%%%%%%%%%%%%%%%%%%%%%%%%%%%%%%%%%%%%%%

\subsection{Model with scalar field}

Let $\phi$ be a scalar field. It is a section of $\ad(\clP)$-valued zero form.
 Consider the action
\beq{act1}
I'(A,\phi,e)=-\frac{e^2}2\int_{\Si_g}d\mu\, (\phi,\phi)-\imath\int_{\Si_g}(\phi,F)\,.
\eq
The gauge group $\clG_U=Map\,(\Si_g\to U)$ acts on $\phi$ and $F$ by the conjugation
\beq{gsa}
\phi\to \Ad_f\phi\,,~~F\to \Ad_fF\,,~~f\in\clG_U\,.
\eq
The partition function
 is defined by the functional integral
\beq{f1}
 Z'( U,\Si_g,q)=\int D\phi DA\exp -\left(\frac{e^2}2\int_{\Si_g}d\mu\, (\phi,\phi))\right)
\cdot \exp-\left((\imath\int_{\Si_g}(\phi,F(A))\right)\,.
\eq
Note that in the limit $e\to 0$ the theory becomes topological. It is a special form
of the $3d$ Chern-Simons theory.

Performing first the Gaussian integral over $\phi$, we come to the partition (\ref{fi})
$$
Z^{II}( U,\Si_g,e)=\int DA\exp\f1{2e^2}\left(\int_{\Si_g}d\mu\, (F,F)\right)\,.
$$

In a similar way we want to define in the the partition function $Z^I$.
To do this we need to describe
the so-called symmetric pairs.

%%%%%%%%%%%%%%%%%%%%%%%%%%%%%%%%%%%%%%%%%%%%%%%%%%%%%%%%%%%%%%%%%%%
\subsection{Type I symmetric spaces}

Here we define the theory that leads to the partition function
$Z^I( U,\Si_g,q)$ (\ref{5}).

Let $G^\mC$ be a simple complex Lie group. The group $G^\mC$ has at least two real
forms. The first is the maximal compact subgroup  $U$ and the second is the normal form  $G^\mR$
\cite{He1}). We give exact definition of the normal form below.
 The subgroups $U$ and $G^\mR$ are fixed point sets of the
commuting anti-holomorphic involutive automorphisms
$\si$ and $\rho$ of $G^\mC$
$$
\rho(U)=U\,,~~\si(G^\mR)=G^\mR\,.
$$
For example, for $G^\mC=\SLN$ the maximal compact subgroup is SU$(N)$
 and $G^\mR$ is SL$(N,\mR)$, $\rho(g)=(g^\dag)^{-1}$, $\si(g)=\bar{g}$.

 Let $K$ be a maximal compact subgroup of $G^\mR$:
 $$
 K=\{g\in G^\mR\,|\,\rho(g)=g\}\,.
 $$
 Or, equivalently,
 $$
 K=\{g\in U\,|\,\si(g)=g\}\,.
 $$
  For the group
SL$(N,\mC)$ the subgroup $K$ is SO$(N,\mR)$.

Let $\gg^\mC=$Lie$(G^\mC)$  be a simple complex Lie algebra and
$\gu=$Lie$(U)$, $\gg=$Lie$(G^\mR)$ its subalgebras.
Let $\ti\rho$ and $\ti\si$
be the involutive automorphisms of $\gg^\mC$ corresponding to $\rho$ and $\si$. Then
\beq{u}
\ti\rho(\gu)=\gu\,,~~\ti\si(\gg)=\gg\,.
\eq
The involutive automorphism $\ti\si$ of $\gg^\mC$
fixes the  real form $\gg$. Then $\ti\rho\cdot\ti\si=\ti\si\cdot\ti\rho$ and $\ti\si$ is an involutive automorphism
of $\gu$ which fix subalgebra $\gk=$Lie$(K)$.
The Lie algebra $\gu$ is the direct sum
\beq{cd}
\gu=\gk\oplus\imath\gp\,,~~\si(\gk)=\gk\,,~~\si(\gp)=-\gp\,.
\eq

Similarly, $\rho$ is an involutive automorphism
of $\gg$ which fix subalgebra $\gk$ and
\beq{rg}
\gg=\gk\oplus\gp\,.
\eq
According with this the Cartan subalgebra $\gh$ is decomposed as a sum of two Cartan subalgebras
\beq{css}
\gh=\gh_{\gk}+\gh_{\gp}\,.
\eq

This $\mZ_2$ grading leads to the commutation relations
\beq{cr}
[\gk,\gk]\subset\gk\,,~~[\gp,\gp]\subset\gk\,,~~
[\gk,\gp]\subset\gp\,.
\eq
For example, for $\gu=su(N)$, $\si(x)=x^T$ and $\gk=so(N)$,
$$
\gp=\{x\in\gu\,|\,x^T=x\}\,.
$$
 The
 subalgebra $\gk$ is orthogonal to $\gp$ with respect to the Killing form on $\gu$.
 \beq{xk}
 (\gk,\gp)=0\,.
 \eq

Let $G$, $U$, $K$ be Lie groups with the Lie algebras $\gg$, $\gu$, $\gk$.
The quotient space $U/K$ is the compact symmetric space and $G/K$ is dual to
it noncompact symmetric space. The compact symmetric spaces $U/K$ are called
the type I symmetric spaces \cite{He1}.
Note that the space $\imath\gp$ is the tangent space to  $U/K$.
\emph{The rank of symmetric space }  $G/K$ (or $U/K$) is
$\dim\,\gh_\gp$.

Here are the examples of type I symmetric spaces $X=U/K$.
For $U=SU(N)$ there are three type I symmetric spaces
$$
1)\, K=SU(p)\times U(q)\,,~p+q=N\,,~~X=U/K~- grasmanians
$$
$$
2)\,K=SO(N)\,,~~X=SU(N)/SO(N)\,,
$$
$$
3) U=SU(2N)\,,~K=Sp(N)\,,~X=SU(2N)/Sp(N)\,.
$$
For $U=SO(N)$
$$
K=SO(p)\times SO(q)\,,~p+q=N,~~X=S^{N-1}=SO(N)/SO(N-1)\,,
$$
and
$$
K=SO(N-1)\,,~~X=S^{N-1}=SO(N)/SO(N-1)\,.
$$
A complete classification of these spaces is presented in Tables 1 and 2.
%%%%%%%%%%%%%%%%%%%%%%%%%%%%%%%%%%%%%%%%%%%%%%%%%%%%%%%%%%%%%%%%%%%%%%%%%%%%%%%%%%%%%%%%

%%%%%%%%%%%%%%%%%%%%%%%%%%%%%%%%%%%%%%%%%%%%%%%%%%%%%%%%%%%%%%%%%%%

\subsubsection{Classification of type I symmetric spaces }
Let $G$ and $K$ be the Lie groups corresponding to the Lie algebras $\gg$ and $\gk$.
The symmetric space $X^-=G/K$ is the Cartan dual
to the compact symmetric space $X=U/K$.

 We describe the spherical representations of
the groups $G$ and $U$. By definition the corresponding modules  have a fixed vectors
invariant under the $K$ actions.
The importance of spherical representations lies, in particular, in the contribution they give to the
 "Fourier transforms" on $X$ and $X^-$.

We defined above (\ref{css}) a maximal abelian subspace $\gh_\gp$ of $\gp$ ($\gh_\gp\subset\gp$. It is the Cartan subalgebra of the
 symmetric space $X$). The Lie algebra $\gg$ has the root decomposition
with respect to the adjoint action of $\gh$
\beq{rd}
\gg=\gg_0 \oplus\sum_{\al\in R_X}\gg_\al\,,~~\ad_\gh\gg_\al=\lan\al,\gh\ran\gg_\al\,,~~
\al\in\gh^*\,.
\eq
Here $R_X=\{\al\}$ is the root system of the symmetric space $X^-$.
%It is the extension
%of the standard root system corresponding to the symmetric space $X^\mC=G^\mC/K$.
The $R_X$ systems have two differences compared to the standard root systems:\\
1) the roots are equipped with multiplicities  $m_\al=\dim\,g_\al$;\\
2) there exists a root system of type BC$_l$ such that there are some roots proportional  to a root $\al$ are
$\pm\al$ and $\pm 2\al$.

The triple $(\gh_\gp,R_X,m)$, where $m\,:\,\al\to m_\al$ is the multiplicity function
defines the real algebra $\gg$ up to isomorphism \cite{He2}.
The Cartan classification of the symmetric spaces  $X=U/K$ is presented Tables
%%%%%%%%%%%%%%%%%%%%%%%%%%%%%%%%%%%%%%%%%%%%%%%%%%%%%%%%%%%%%%%%%%%%%%%%%%%%%%%%%%%%%%%

\begin{center}
\begin{tabular}{|c|c|c|c|c|c|
 }
  \hline
  \hline
  % after \\: \hline or \cline{col1-col2} \cline{col3-col4} ...

  Type &$U/K$ & rank & Root\,system& Multiplicities \\
 \hline
 \hline
 AI &$SU(N)/SO(N)$ &N-1 & $A_{N-1}$ & $m_\al=1$  \\
 \hline
 AII &$SU(2N)/Sp(N)$ &N-1 & $A_{N-1}$ & $m_\al=4$   \\
 \hline
 AIII & $
 SU(p+q)/S(U(p)\times U(q))$ &$ \begin{array}{c}
                                n= \\
                                 min(p,q)
                              \end{array}$
  &$\begin{array}{cc}
p=q & C_n \\
p>q & BC_n\\
 &
\end{array}$
 &
$\begin{array}{ll}
  m_\al=2 & m_{2\be}=1 \\
  m_\al=2  & m_\be=2(p-q)\\
  & m_{2\be}=1
\end{array}  $
\\
\hline
BDI* &
$
 SO(p+q)/SO(p)\times SO(q))$ &$ \begin{array}{c}
                                n= \\
                                 min(p,q)
                              \end{array}$
  &$\begin{array}{cc}
p=q & D_n \\
p>q & B_n
\end{array}$
 &
$\begin{array}{ll}
  m_\al=1 & \\
  m_\al=1  & m_\be=(p-q)
\end{array}  $
\\
\hline
DIII & SO(2n)/U(n) & [n/2]
 &$\begin{array}{cc}
n=2\nu & C_\nu \\
n=2\nu+1 & BC_\nu
\end{array}$
 &
$\begin{array}{ll}
  m_\al=4 & m_{2\be}=1 \\
  m_\al=4  & m_\be=4,\,m_{2\be}=1
\end{array}  $
\\
\hline
CI* & $Sp(n,\mR)/U(n)$ &  $n $ & $C_n$ & $m_\al=1$ \\
\hline
CII & $Sp(p+q)/Sp(p)\times Sp(q)$ &
$ \begin{array}{c}
                                n= \\
                                 min(p,q)
                              \end{array}$
 &$\begin{array}{cc}
p=q & C_n \\
p>q & BC_n\\
  &
\end{array}$
 &
$\begin{array}{ll}
  m_\al=4 & m_{2\be}=3 \\
  m_\al=4  & m_\be=4(p-q)\\
  & m_{2\be}=3
\end{array}  $
\\
\hline
\hline
\end{tabular}

\bigskip
\textbf{Table 1. Classical type I symmetric spaces}

\bigskip

\begin{tabular}{|c|c|c|c|c|c|
 }
  \hline
  \hline
  % after \\: \hline or \cline{col1-col2} \cline{col3-col4} ...

  Type &$U/K$ & rank & Root\,system& Multiplicities \\
 \hline
 \hline
 EI*& $E_6/Sp(4)$ & 6 & $E_6$ & $m_\al=1$ \\
 \hline
 EII & $E_6/SU(6)\times SU(2)$ & 4 & $F_4$ & $m_\al=(1,1,2,2)$\\
 \hline
EIII & $E_6/SO(10)\times SO(2)$ &2&  $BC_2$ & $m_\al=6,m_\be=8,m_{2\be}=1$\\
 \hline
EIV & $E_6/F_4$ &2&  $A_2$ &$m_\al=8$\\
 \hline
 EV* & $ E_7/SU(8)$ &7& $E_7$ & $m_\al=1$  \\
\hline
 EVI & $ E_7/SO(12)\times SU(2)$ &4& $F_4$ & $m_\al=(1,1,4,4)$\\
 \hline
 EVII & $ E_7/E_6\times SO(2)$ &3& $C_3$ &  \\
 \hline
 EVIII* & $ E_8/SO(16)$ & 8& $E_8$ & $m_\al=1$ \\
 \hline
 EIX & $E_8/E_7\times SU(2)$ &4 & $F_4$  & $m_\al=(1,1,8,8)$\\
 \hline
 FI* &$F_4/Sp(3)\times SU(2)$ & 4& $F_4$ & $m_\al=1$\\
 \hline
 FII &$F_4/SO(9)$&1& $BC_1$& $m_\be=8, m_{2\be}=7$
 \\
 \hline
 GI* & $G_2/SU(2)\times SU(2)$ &2&$G_2$& $m_\al=1$\\
 \hline
 \hline
 \end{tabular}

 \bigskip
\textbf{Table 2. Exceptional types I symmetric spaces}

\end{center}

Let $R_X^+$ be the set of positive roots with respect to some ordering
($R_X=R_X^+\cup -R_X^+$). The space
\beq{ns}
\gn=\sum_{\al\in R_X^+}\in\gg_\al
\eq
is the nilpotent subalgebra. The algebra $\gg$ is
the direct sum of its subalgebras (the Iwasawa decomposition)
\beq{ida}
\gg=\gk+\gh+\gn\,.
\eq
% where the compact subalgebra $\gu$ is defined by the involutive automorphism (\ref{cd}).

%%%%%%%%%%%%%%%%%%%%%%%%%%%%%%%%%%%%%%%%%%%%%%%%%%%%%%%%%%%%%%%%%%%%%%%%%%%%%%%%%%%%%%%%%%%%

 \subsection{Gauge theory }

The gauge fields $\phi$ and $F$ in the action (\ref{act1}) take values in the
compact Lie algebra $\gu=$Lie$(U)$.
According with (\ref{cd})
$$
\phi=\phi_\gk+\phi_\gp\,,~~\phi_\gk\in\gk\,,~\phi_\gp\in\imath\gp\,.
$$
Consider the gauge action (\ref{gsa})
on the field $\phi$.  One can make  the $\gk$ component $\phi_\gk$
vanish
\beq{sb}
\phi_\gk=0\,.
\eq
This partial gauge fixing  breaks the gauge group $\clG_U$ to the subgroup
$\clG_K=Map\,(\Si_g\to K)$.

Consider the second term $\int_{\Si_g}(\phi,F)$
in the action (\ref{act1}).
 After the gauge fixing  $F(A)$ becomes defined
up to the adding an element from $\gk$. In other words, it is an element of
the quotient space $F(A)\in\gu/\gk\sim\imath\gp$. It is the tangent space to
the symmetric space $U/K$.

Consider a disc $D\subset\Si_g$ with the circle $S$ as its boundary.
% such that the value of the integral
%$\int_DF(A)$ is a nontrivial element in the tangent space $\imath\gp$.
Because $\int_DF(A)=\oint_S A$ the holonomy is an element of the symmetric space $U/K$
$$
\clU_S=P\exp\oint_S A\in U/K\,.
$$

In this way
the trace of the holonomy $\tr\,\clU_w$ in the lattice approximation
 (\ref{mp}) is a $K$-invariant function on the symmetric space $U/K$, or in other words, is a function on the double coset space $K\setminus U/K$.
   This type of functions can be expanded in the basis of zonal spherical
 functions $\varphi_\nu(u)$, $\nu\in\G(U/K)$ defined in next subsection.

 %We can look at this construction from a slightly different perspective.
%  Define the associated bundle
%  $E_\nu=\clP\otimes_{G^\mC}V_\nu$, where $\nu\in\G(U/K)$. Choose $\clG_K$ invariant
%  section $s\in\G(E_\nu)$. This choice break the gauge group $\clG_U\to\clG_K$.

%%%%%%%%%%%%%%%%%%%%%%%%%%%%%%%%%%%%%%%%%%%%%%%%%%%%%%%%%%%%%%%%%%%%%%%%%%%%%%%%%%%%%%%

\subsubsection{The zonal spherical functions }
 The zonal spherical functions (ZSF) are defined as follows.
 Let $\pi_\nu$ be a class one representation in
 the space $V_\nu$, and $\{e_j\}$, $j=1,\dots,d_\nu$ is an orthonormal basis in   $V_\nu$
 with respect to the Hermitian $U$-invariant form in $V_\nu$, $\,((e_j,e_k)=\de_{jk})$.
 The matrix elements $t_{jk}^\nu(u)=(\pi_\nu(u)e_j,e_k)$ form a complete orthogonal
 system in the space of functions on the group $U$ \cite{Vi} ch.I, $\S 4$, Th.1.
 Let $du$ be the normalized Haar measure on $U$. Then
 \beq{ora}
 \int_Udu\,t_{jk}^\nu(u)\bar t_{mn}^\mu(u)du=\frac{\de_{\mu\nu}\de_{jm}\de_{nk}}{d_\nu}\,.
 \eq

 Choose the basis such that $e_1$ is the $K$-invariant vector. Then ZSF is defined as the special matrix element
 \beq{sf}
 \varphi^U_\nu(u):=t^\nu_{11}=(\pi_\nu(u) e_1,e_1)\,,~~\nu\in\G(U/K)\,.
 \eq
 It follows from this definition that:\\
 \begin{subequations}\label{gtr}
  \begin{align}
 & 1. ~\varphi^U_\nu(Id)=1\,,\\
 &2. ~\varphi^U_\nu(k_1uk_2)=\varphi^U_\nu(u)\,,~k_j\in K\,.
 \end{align}
  \end{subequations}
  Due to the unitarity of representations
  % $\overline{\pi_\nu(u^{-1})}$
  \beq{unz}
\overline{\varphi^U_\nu(u^{-1})}=\varphi^U_\nu(u)\,.
\eq
and
\beq{com}
\varphi^U_\nu(u_1u_2)=\varphi^U_\nu(u_2u_1)\,.
\eq
  ZSF form an orthogonal basis in the space of $K$-invariant functions on $X=U/K$ (see (\ref{ora})).
 \beq{o10}
\int_U\varphi^U_\nu(u)\bar\varphi^U_\mu(u)du=\frac{\de_{\mu\nu}}{d_\nu}\,.
\eq
In particular,  the delta functional
$\int_Uduf(u)\de(u)=f(Id)$ is defined on the  the double coset space $K\setminus U/K$ and in this way can be expanded in the ZSF series. This expansion takes the form
\beq{de}
\de(u)=\sum_{\nu\in\G^+(U/K)}d^I_{\nu}\varphi^U_\nu(u)\,.
\eq
ZSF are the eigen-functions of the radial part of the second Casimir operator $c_2$. The latter coincides
with the radial part of the Laplace-Beltrami operator on  $X=U/K$.

 Note that there are ZSF $\varphi^G_\nu(x)$ ($x\in G/K$) related to the dual symmetric space $G/K$ defined in Sect.3.2.
From now on, we will omit the  sign $U$ or $G$ in $\varphi^{U,G}_\nu$ when it is clear what we are talking about.

  More generally, ZSF form the following algebra with respect to the convolution
 \beq{o2a}
\int_U\varphi_\nu(u)\varphi_\mu(u^{-1}v)du=\frac{\de_{\mu\nu}}{d_\nu}\varphi_\nu(v)
\eq

and
\beq{o3}
\int_Udv\varphi_\nu(Av)\varphi_\mu(v^{-1}B)=\frac{\de_{\mu\nu}}{d_\nu}\varphi_\nu(AB)\,.
\eq
The proof of (\ref{o10}), (\ref{o2a}), (\ref{o3}) is based on the definition of ZSF (\ref{sf})
and on the orthogonality relation (\ref{ora}).

Consider the characters for the class one representations $\nu\in\G(U/K)$
 \beq{chi}
 \chi_\nu=\sum_{j=1}^{d_\nu} t_{jj}^\nu(u)\,.
 \eq
It follows from (\ref{ora}) that characters  form the orthonormal basis in the space of functions on $U$-invariant under the conjugation
\beq{oc}
\int_Udu\chi_\nu\bar\chi_\mu=\de_{\mu\nu}\,.
\eq
Using the definition (\ref{chi}) we find that the characters satisfy (\ref{f5}) and (\ref{f51}).
%(compare with (\ref{o1})) and
%\beq{f2}
% \int_U\chi_\nu(uAu^{-1}B)du=\f1{d_\nu}\chi_\nu(A)\chi_\nu(B)\,.
%\eq
There are the following interrelations between characters and ZSF
 \beq{o2}
\int_U\chi_\nu(u)\varphi_\nu(u^{-1}v)du=\frac{1}{d_\nu}\varphi_\nu(v)\,.
\eq
\beq{f3}
 \int_U\varphi_\nu(uBu^{-1})du=\f1{d_\nu}\chi_\nu(B)\varphi_\nu(A)\,,~~
%\eq
%\beq{f3}
 \int_U\varphi_\nu(uAu^{-1}B)du=\f1{d_\nu}\chi_\nu(A)\varphi_\nu(B)\,.
\eq
These equalities  follow from (\ref{ora}),
(\ref{sf}), (\ref{gtr}) and (\ref{chi}).

  As an example, the ZSF for the sphere $S^2=SO(3)/SO(2)$ are the Legendre polynomials.
 For the spaces of type II (compact groups) ZSF are proportional to the characters
 \beq{czf}
\varphi_\nu(u)=\chi_\nu(u)/d_\nu\,.
 \eq

Since after the gauge fixing the trace of holonomy $\tr\,\clU_w$ becomes an invariant function on the type I symmetric space
$U/K$ it can be expanded in the basis of the ZSF. In this way we replace (\ref{mp}) by the expression
\beq{mp1}
\mathcal{G}^I_w(\clU_w,\varrho_w)=\sum_{\nu\in  \G^+(U/K)}d_\nu\varphi_\nu(u)\exp(-\varrho_w c_2(\nu)/2)\,,
\eq
In the continuum limit $\varrho_w\to 0$ the holonomy around plaquette $w$ becomes trivial $\clU_w\to Id$.  It follows  from (\ref{de}) that
\beq{cl}
\lim_{\varrho_w\to 0}\mathcal{G}^I_w(\clU_w,\varrho_w)=\de(\clU_w-Id)\,.
\eq
The partition function in the lattice form is (compare with (\ref{fa}))
\beq{fa3}
Z^{I}(U,\Si_g)=
\int \prod_\ga du_\ga\prod_i\mathcal{G}_{w_i}^I(u_i,\varrho_i)\,,
\eq
where $\mathcal{G}_{w_i}^I$ is (\ref{mp1}).
%\beq{fa4}
%\G(u_i)=\sum_{\nu\in  \G^+(U/K)}d_\nu\varphi_\nu(u_i)\exp(-\varrho_{i} c_2(\nu)/2)\,,
%\eq
%(see  (\ref{mp}), (\ref{fa})).

The invariance under the subdivisions of lattices follows from (\ref{o3}).
We use the same argumentation as in \cite{Wi} and replace the characters by the ZSF.
The partition function (\ref{fa1})  becomes
\beq{z1}
Z^I( U,\Si_g,q)=\sum_{\nu\in  \G^+(U/K)}d_\nu e^{-\rho c_2(\nu)/2}
\int du_idv_j\varphi_\nu(u_1v_1u_1^{-1}v_1^{-1}\cdots u_gv_gu_g^{-1}v_g^{-1})\,.
\eq
From (\ref{f3}) every time one integrates over $u_i$ or $v_j$ one gets a factor of $1/d_\nu$.
On the last step we come to the integral
$$
\f1{d^{2g-1}_nu}\int_Udu_g\varphi_\nu(u_g)\chi_\nu(u_g^{-1})
\stackrel{(\ref{o2})}{=}\frac{\varphi_\nu(Id)}{d^{2g}_\nu}=\f1{d^{2g}_\nu}\,.
$$
Using this expression we come to the partition function (\ref{5}) and eventually to
the type I Witten zeta-function (\ref{wi1}).
%\beq{z2}
%\zeta^{I}_U(s)=\sum_{\nu\in  \G^+(U/K)}\frac{1}{d^{s}_\nu}\,.
%\eq
%\beq{z2}
%Z^I( U,\Si_g,q)=\sum_{\nu\in  \G^+(U/K)}\frac{q^{c_2(\nu)}}{d^{2g-1}_\nu}\,.
%\eq
%%%%%%%%%%%%%%%%%%%%%%%%%%%%%%%%%%%%%%%%%%%%%%%%%%%%%%%%%%%%%%%%%%%%%
%%%%%%%%%%%%%%%%%%%%%%%%%%%%%%%%%%%%%%%%%%%%%%%%%%%%%%%%%%%%%%%%%
\subsubsection{Twisted External States}

Consider a surface $\Si_g$
with boundary consisting of a circle $S$.
We want to calculate the Yang-Mills path integral over all connections
on $\Si_g$ with the holonomy $u$ around $S$.
 %for which the holonomy $\clU_S(\Theta)$ around $S$ in the conjugacy class, defined
%by an element $\Theta\in\imath\gp$, where $\imath\gp$ is the
%Cartan subalgebra of the symmetric space $U/K$ (\ref{cd}).
% The functional integral over all connections
% for the fixed value of the holonomy around $S$ defines a function of $\psi(\clU_S)$.
Using (\ref{z1}) we obtain
$$
\psi(u)=\sum_{\nu\in  \G^+(U/K)}d_\nu e^{-\rho c_2(\nu)/2}
\int dXdu_idv_j\varphi_\nu(XuX^{-1} u_1v_1u_1^{-1}v_1^{-1}\cdots u_gv_gu_g^{-1}v_g^{-1})\,.
$$
To calculate $\psi(u)$, we use the same procedure as for calculating the function $Z^I$. Then
we obtain
\beq{ps}
\psi(u)=\sum_{\nu\in  \G^+(U/K)} \frac{e^{-\rho c_2(\nu)/2}\chi_\nu(u)}{d_\nu^{2g}}\,.
\eq

More general, for a surface   $\Si_{g,n}$ with $n$ holes with the holonomies $u_i$
we have
\beq{ps1}
\psi(u_1,\ldots,u_n)=
\sum_{\nu\in  \G^+(U/K)} \frac{e^{-\rho c_2(\nu)/2}\prod_{i=1}^n\chi_\nu(u_i)}{d_\nu^{2g+n-1}}\,.
\eq

%\beq{z1}
%Z^I( U,\Si_g,q)=\sum_{\nu\in  \G^+(U/K)}d_\nu e^{-\rho c_2(\nu)/2}
%\int du_idv_j\varphi_\nu(u_1v_1u_1^{-1}v_1^{-1}\cdots u_gv_gu_g^{-1}v_g^{-1})\,.
%\eq
%%%%%%%%%%%%%%%%%%%%%%%%%%%%%%%%%%%%%%%%%%%%%%%%%%%%%%%%%%%%%%%%%%%%%%
%%%%%%%%%%%%%%%%%%%%%%%%%%%%%%%%%%%%%%%%%%%%%%%%%%%%%%%%%%%%%%%%%%

\subsection{Zeta-functions and Harisch-Chandra c-function}

The dimensions of the class one representations (\ref{wi1}) is defined in
terms of the  Harisch-Chandra c-function  \cite{He2,V}.

Let $N$ be the maximal nilpotent subgroup of the real group $G$ corresponding to the nilpotent
subalgebra $\gn$ (\ref{ns}) and $H$ is the Cartan subgroup $(H,N\subset G)$
($Lie(H)=\gh$ in (\ref{ida})).
The Iwasawa decomposition of the group $G$ corresponding to (\ref{ida})
has the form
$$
G=KHN\,.
$$
%Here $H$ is the Cartan subgroup of the symmetric case $G/K$.
%Then $K\cap H=\oslash$.
Let $h(g)$ be the Cartan part of the decomposition $g=kh(g)n$
and $\la\in\gh^*$.
The zonal spherical functions on the noncompact symmetric space $X^-=G/K$
 has the integral representation \cite{HC}
$$
\phi^G_\la(g)=\int_Kh^{\imath\la-\rho_X}(gk)dk\,,~~
h^{\imath\la-\rho_X}(gk)=\exp\lan\imath\la-\rho_X,\log h(gk)\,,
$$
where $dk$ is normalised measure on the group $K$ and
\beq{ro}
\rho_X=\oh\sum_{\al\in R_{X}^+}(m_\al\al+2m_{2\al}\al)\,.
\eq
 The Harisch-Chandra c-function
is defined as the limit of the spherical functions
$$
c(\la)=\lim_{t\to-\infty}e^{\lan-\imath\la+\rho_X,t\xi\ran}
\phi^G_\la(\exp\,t\xi)\,.
$$
It has the integral representation over $N$
$$
c(\la)=\int_Nh^{\imath\la-\rho_X}(n)dn\,,~~\la\in\gh^*\,.
$$

Let $R_{X,0}^+$ be the subset of positive roots that are not integer multiples
other positive roots.
It was proved in \cite{GK} that
\beq{gk1}
c(\la)=\prod_{\al\in R_{X,0}^+}c_\al(\la)\,,
\eq
and
\beq{gk2}
c_\al(\la)=\frac{2^{-\imath(\la,\al_0)}\G(\imath(\la,\al_0))}{\G(\frac{m_\al}4+\oh+ \oh(\imath\la,\al_0))\G(\frac{m_\al}4+\frac{m_{2\al}}2+ \oh(\imath\la,\al_0))}\,,~~
\al_0=\frac{\al}{(\al,\al)}\,.
\eq

\section{Representations of class one}
%Compact symmetric spaces}

Consider the irreducible representations appearing in the decomposition of the unitary representation of $U$ in the space $L^2(U/K)$.
 They are precisely defined above the class one representations. The $\acute{E}.$ Cartan-Helgason theorem
\cite{Wa} asserts that the class one representations are defined by the
class one dominant weights
\beq{co}
\G^+(U/K)=\{\nu\in\gh^*\,|\,1.\,\nu|_{\gk^*}=0\,,~2.\,(\nu,\al)/(\al,\al)\in\mZ^+,~\al\in R_X^+\}\,.
\eq
%More concretely,  the function $f(u)\in L^2(K\setminus U/K)$ can be expanded into a series
%of ZSF
%$$
%f(u)=\sum_{\nu\in\G^+(U/K)}c_\nu\varphi_\nu(u)\,,
%~~c_\nu=d_\nu\int_Uduf(u)\varphi_\nu(u)\,,~~d_\nu=\dim\,V_\nu\,.
%$$

The dimension $d_\nu$ for $\nu\in\G^+(U/K)$ is related to the Harisch-Chandra function \cite{He2,V}
\beq{dim1}
d_\la=\prod_{\al\in R_{X,0}^+}d_\al(\la)\,,~~
d_\al(\la)=\frac{c_\al(\imath\rho_X)}{c_\al(\imath(\la+\rho_X))}\cdot
\frac{c_\al(-\imath\rho_X)}{c_\al(-\imath(\la+\rho_X))}
\eq

Let $\ti R_X$ be a subsystem of roots $\ti R_X=\{\al,\,2\al\in R_X\}$.
For these roots condition 2 in (\ref{co}) is modified as
\beq{iw1}
\frac{(\la,\al)}{(\al,\al)}\in 2\mZ\,,\,\al\in \ti R_X\,.
\eq

Let $\al_0=\al/(\al,\al)$. Introduce notations
\beq{rl1}
\rho_{X\al}=(\rho_X,\al_0)=\sum_{\be\in R^+_X}(\oh m_\be+m_{2\be})\frac{(\be,\al)}{(\al,\al)}\,,
\eq
\beq{rl}
(\la,\rho_X)_\al=(\la+\rho_X,\al_0)=n_\al +\rho_{X\al}\,,~~n_\al\in\mZ^+
%\la\in\G^+(U/K)\,,
\eq
where $n_\al$ is even in the case (\ref{iw1}).
 Then (\ref{dim1}) takes the form
$$
d_\al(n_\al)=
\frac{\G(-\rho_{X\al})\G(\rho_{X\al})}
{\G(\frac{m_\al}4+\oh-\frac{\rho_{X\al}}2)\G(\frac{m_\al}4+\frac{m_{2\al}}2-\frac{\rho_{X\al}}2)
\G(\frac{m_\al}4+\oh+ \frac{\rho_{X\al}}2)\G(\frac{m_\al}4+\frac{m_{2\al}}2+ \frac{\rho_{X\al}}2)}
\times
$$
$$
\frac{\G(\frac{m_\al}4+\oh- \frac{(\la,\rho_X)_\al}2)
\G(\frac{m_\al}4+\frac{m_{2\al}}2 -\frac{(\la,\rho_X)_\al}2)
\G(\frac{m_\al}4+\oh+ \frac{(\la,\rho_X)_\al}2)
\G(\frac{m_\al}4+\frac{m_{2\al}}2+ \frac{(\la,\rho_X)_\al}2)}{
\G(-(\la,\rho_X)_\al)\G((\la,\rho_X)_\al)}\,.
$$
Rewrite it as
\beq{mr}
d_\al(n_\al)=D_{X,\al}\times
\eq
$$
\frac{\G(\frac{m_\al}4+\oh- \frac{(\la,\rho_X)_\al}2)
\G(\frac{m_\al}4+\frac{m_{2\al}}2 -\frac{(\la,\rho_X)_\al}2)
\G(\frac{m_\al}4+\oh+ \frac{(\la,\rho_X)_\al}2)
\G(\frac{m_\al}4+\frac{m_{2\al}}2+ \frac{(\la,\rho_X)_\al}2)}{
\G(-(\la,\rho_X)_\al)\G((\la,\rho_X)_\al)}\,,
$$
$D_{X,\al}$ is given by $d_\al(0)=1$.

Substituting it in (\ref{wi1}) we come to the type I zeta functions.
Our goal is to simplify this expression.
%%%%%%%%%%%%%%%%%%%%%%%%%%%%%%%%%%%%%%%%%%%%%%%%%%%%%%%%%%%%%%%%%%%%%%%%%%%%%%%%%%

\subsubsection{Symmetric spaces with $m_{2\al}=0$}
Assume that $m_{2\al}=0$.
Using the
Legendre duplication formula
\beq{ld}
\G(2z)=2^{2z-1}\pi^{-\oh}\G(z)\G\left(z+\oh\right)\,.
\eq
and taking $z=\oh(m_\al/2+(\imath\la,\al_0))$ rewrite (\ref{gk2})
\beq{gk3}
c_\al(\la)=\frac{2^{\oh m_\al-1}\G(\imath\la,\al_0)}{\pi^\oh\G(\oh m_\al+(\imath\la,\al_0))}\,.
\eq
Taking into account (\ref{rl})
%($n_\al=(\rho_{X\al}-\imath(\la,\rho_X)_\al)$).
rewrite $d_\al(n_\al)$ (\ref{mr})
\beq{di1}
d_\al(n_\al)=
D_{X,\al}
\frac{\G(\oh m_\al+n_\al+(\rho_X)_\al)\G(\oh m_\al-n_\al-(\rho_X)_\al)}
{\G(n_\al+(\rho_X)_\al)\G(-n_\al-(\rho_X)_\al)}\,.
\eq
$$
D_{X,\al}=
\frac{\G(-\rho_{X\al})\G((\rho_{X\al})}
{\G(\frac{m_\al}2-\rho_{X\al})
\G(\frac{m_\al}2+ \rho_{X\al})}
$$

For the normal symmetric spaces $m_\al=1$, $R_X=R$,
$$
(\rho_X)_\al=\oh\sum_{\be\in R^+}\frac{(\be,\al)}{(\al,\al)}\,.
$$
  Since
\beq{tgf}
\G(x)\G(-x)=\frac{\pi}{x\sin\pi x}\,, ~~
\G(\oh+x)\G(\oh-x)=\frac{\pi}{\cos\pi x}\,
\eq
we have
\beq{gs}
\frac{\G(\oh+ z)\G(\oh- z)}{\G( z)\G(- z)}=
\frac{z\sin\pi z}{\cos\pi z}\,.
\eq
Thus, if $X$ - \emph{normal symmetric space} then
\beq{di2}
d_\al(n_\al)=\frac{\cos\pi (\rho_X)_\al}{(\rho_X)_\al\sin\pi (\rho_X)_\al}
\frac{(n_\al+(\rho_X)_\al)\sin\pi (n_\al+(\rho_X)_\al}{\cos\pi (n_\al+(\rho_X)_\al)}=
\frac{n_\al+(\rho_X)_\al}{(\rho_X)_\al}\,.
\eq

For \emph{$X$ being compact groups} (the symmetric spaces of type II) $m_\al=2$.
 From (\ref{gs}) $d_\al(n_\al)$ (\ref{di1}) is equal
\beq{di6}
d_\al(n_\al)=
\left(\frac{n_\al+(\rho_X)_\al}{ (\rho_X)_\al}\right)^2\,.
\eq

For arbitrary values of multiplicities we use the equality
\beq{gs1}
\G(1+n+x)=(n+x)(n-1+x)\cdots x\G(x)\,.
\eq
Define two types of polynomials of degree $2l$
\beq{c0}
c_0(x,l)=\frac{\G(l+x)\G(l-x)}{\G(x)\G(-x)}=(-1)^l(x-l+1)(x-l+2)\cdots x^2(x+1)\cdots(x+l-1)\,.
\eq
Note that $c_0(0,l)=(-1)^{l}(2l-1)!l$.

To define the polynomial $c_1(x,l)$ consider the function
\beq{om}
\frac{\G(l+\oh+x)\G(l+\oh-x))}{\G(x)\G(-x))}=c_1(x,l)\frac{\G(\oh+x)\G(\oh-x))}{\G(x)\G(-x))}
%{\G(n_\al+(\rho_X)_\al)\G(-n_\al-(\rho_X)_\al)}=
=c_1(x,l)\frac{x\sin\pi x}{\cos\pi x}\,.
\eq
Hear
\beq{c1a}
c_1(x,l)=(-1)^l(x+\oh -l)\cdots(x-\oh)(x+\oh)\cdots(x-\oh+l)=
\eq
$$
\frac{(-1)^l}{2^{2l}}(2x+1 -2l)\cdots(2x-1)(2x+1)\cdots(2x-1+2l)
$$
 and $c_1(0,l)=(-1)^l(l^2-\f1{4})\cdots(l^2-(l^2-\left(\frac{l-1}2\right)^2)$.

We need the following identities for gamma-functions
\beq{gs7}
\G(1+n+x)\G(1+n-x)=\frac{\pi x}{\sin\pi x}\prod_{k=1}^n(k^2-x^2)=
\frac{\pi x}{\sin\pi x}c_0(x,n+1)\,,~for~n\geq1\,,
\eq
\beq{gs2}
\G(\oh+n+x)\G(\oh+n-x)=\frac{\pi}{\cos\pi x}\prod_{k=1}^n((k-\oh)^2-x^2)
=\frac{\pi}{\cos\pi x}c_1(x,n)\,,~for~n\geq1\,.
\eq
%   Then using (\ref{tgf}) we obtain
%\beq{100}
%\frac{\G(1+n+x)\G(1+n-x)}{\G(x)\G(-x)}
%\eq

There are two different classes of formulae for even and odd multiplicities $m_\al$.\\
$\bullet$ \emph{$m_\al=2l_\al$, $l_\al\geq 1$.} \\
Then from (\ref{di1}) and (\ref{c0}) we have
$$
d_\al(n_\al)=
D_{X,\al}
\frac{\G(l_\al+n_\al+(\rho_X)_\al)\G(l_\al-n_\al-(\rho_X)_\al)}
{\G(n_\al+(\rho_X)_\al)\G(-n_\al-(\rho_X)_\al)}=
D_{X,\al}c_0(n_\al+(\rho_X)_\al,l_\al)=\,.
$$
Thus
\beq{di4}
d_\al(n_\al)=\frac{n_\al+(\rho_X)_\al}{(\rho_X)_\al}
\frac{(n_\al+(\rho_X)_\al-l_\al+1)\cdots (n_\al+(\rho_X)_\al)\cdots(n_\al+(\rho_X)_\al+l-1)}
{((\rho_X)_\al-l_\al+1)\cdots(\rho_X)_\al \cdots((\rho_X)_\al+l_\al-1)}
%\frac{\G(l_\al+n_\al+(\rho_X)_\al)\G(l_\al-n_\al-(\rho_X)_\al)}
%{\G(n_\al+(\rho_X)_\al)\G(-n_\al-(\rho_X)_\al)}=
%D_{X,\al}c_0(n_\al+(\rho_X)_\al,l_\al)=\,.
%}{c_0((\rho_X)_\al,l_\al}
\eq
%$$
%\frac{(n_\al+\rho_{X\al})}{\rho_{X\al}
%
%\prod_{k=0}^{l_\al-1}\frac{k^2-(n_\al+\rho_{X\al})^2}{k^2-\rho_{X\al}^2}\,.
%$$

%From (\ref{di4})
%\beq{di10}
%d_\al(n_\al)=\frac{(n_\al+\rho_{X\al})}{\rho_{X\al}}\frac{c_0(n_\al,l_\al)}{c_0(0,l_\al)}=
%\cdot\frac{(n_\al+\rho_{X\al})}{\rho_{X\al}}
%\frac{(n_\al+1)\ldots(n_\al+l_\al)^2\ldots(n_\al+2l_\al-1)}{(2l_\al-1)!}
%\,.
%\eq

$\bullet\bullet$ $m_\al=2l_\al+1$, \\
For $l_\al=0$ we have (\ref{di2}).
 %from (\ref{gs})
%\beq{l0}
%d_\al(n_\al)=\frac{n_\al+(\rho_X)_\al}{(\rho_X)_\al}\,.
%\eq
For $l_\al\geq 1$ we have from (\ref{di1}) and (\ref{om})
$$
d_\al(n_\al)=
D_{X,\al}
\frac{\G(l_\al+\oh+n_\al+(\rho_X)_\al)\G(l_\al+\oh-n_\al-(\rho_X)_\al)}
{\G(n_\al+(\rho_X)_\al)\G(-n_\al-(\rho_X)_\al)}=
$$
$$
D_{X,\al}c_1(n_\al+(\rho_X)_\al,l_\al)
\frac{(n_\al+(\rho_X)_\al)\sin\pi (n_\al+(\rho_X)_\al)}{\cos\pi (n_\al+(\rho_X)_\al)}=
$$
$$
\frac{c_1(n_\al+(\rho_X)_\al,l_\al)}{c_1((\rho_X)_\al,l_\al)}\frac{n_\al+(\rho_X)_\al}{(\rho_X)_\al}
\frac{\sin\pi (n_\al+(\rho_X)_\al)\cos\pi ((\rho_X)_\al)}{\cos\pi (n_\al+(\rho_X)_\al)\sin\pi (\rho_X)_\al}=
$$
$$
\frac{c_1(n_\al+(\rho_X)_\al,l_\al)}{c_1((\rho_X)_\al,l_\al)}\frac{n_\al+(\rho_X)_\al}{(\rho_X)_\al}\,.
$$
From (\ref{c1a})
\beq{di5}
d_\al(n_\al)=\frac{n_\al+(\rho_X)_\al}{(\rho_X)_\al}\times
\eq
$$
\frac{(n_\al+(\rho_X)_\al+\oh -l_\al)\cdots(n_\al+(\rho_X)_\al-\oh)(n_\al+(\rho_X)_\al+\oh)\cdots(n_\al+(\rho_X)_\al-\oh+l_\al))}
{((\rho_X)_\al+\oh -l_\al)\cdots((\rho_X)_\al-\oh)((\rho_X)_\al+\oh)\cdots((\rho_X)_\al-\oh+l_\al))}
$$
%\beq{di5}
%\frac{n_\al+\rho_{X\al}}{\rho_{X\al}}
%\prod_{k=1}^{l_\al}\frac{(k-\oh)^2-(n_\al+\rho_{X\al})^2}{(k-\oh)^2-(\rho_{X\al})^2}\,.
%\eq

%%%%%%%%%%%%%%%%%%%%%%%%%%%%%%%%%%%%%%%%%%%%%%%%%%%%%%%%%%%%%%%%%%%%%%%

\subsubsection{Symmetric spaces with $m_{2\al}\neq0$}

%Note that (see (\ref{gs})) that
%\beq{20}
%\G(-\rho_{X\al})\G(\rho_{X\al})=\frac{\pi}{\rho_{X\al}\sin\pi \rho_{X\al}}
%\eq
%We restrict here the case $m_{2\al}=1$. In this case $m_\al$ is even.

If $m_{2\al}\neq 0$ then $m_{2\al}$ are odd
\beq{m2}
m_{2\al}=1+2j_\al\,,~j_\al=0,1,3\,,
\eq
and $m_\al$ are even (see Table 1). Let
\beq{rj}
m_\al=\left\{\begin{array}{cc}
              A) & 4l_\al\,,~l_\al\in{\mathbb N} \\
              B) & 4l_\al+2 \,,~l_\al\in{\mathbb N}
            \end{array}
\right.~~r_\al=l_\al+j_\al\,.
\eq

Due to (\ref{iw1}) for (\ref{rl}) we have
$(\la,\rho_X)_\al=(\la+\rho_X,\al)=2n_\al +\rho_{X\al}$

$\bullet$ \emph{The case A}.\\
From the general formula (\ref{mr}) and using (\ref{gs2}) we obtain
$$
\G(\frac{m_\al}4+\oh -\oh(2n_\al+\rho_{X\al}))\G(\frac{m_\al}4+\oh +\oh(2n_\al+\rho_{X\al}))\stackrel{(\ref{gs2})}=
\frac{\pi}{\cos\frac{\pi}2(2n_\al+ \rho_{X\al})}c_1(2n_\al+\rho_{X\al},l_\al)\,,
%\prod_{k=1}^{l_\al}((k-\oh)^2-\frac{1}4(\rho_{X\al})^2)
$$
$$
\G(\frac{m_\al}4+\frac{m_{2\al}}2 -\oh(2n_\al+\rho_{X\al}))\G(\frac{m_\al}4+\frac{m_{2\al}}2 +\oh(2n_\al+\rho_{X\al}))\stackrel{(\ref{gs2})}=
\frac{\pi}{\cos\frac{\pi}2(2n_\al+ \rho_{X\al})}c_1(2n_\al+\rho_{X\al},r_\al)\,.
$$
$$
\f1{\G(2n_\al+\rho_{X\al})\G(-2n_\al-\rho_{X\al})}\stackrel{(\ref{tgf})}=
\f1{\pi}(2n_\al+\rho_{X\al})\sin\pi(2n_\al+\rho_{X\al})
$$
From (\ref{mr}) and (\ref{tgf}) we obtain
$$
d_\al(n_\al)=\frac{2n_\al+\rho_{X\al}}{\rho_{X\al}}
\frac{\sin\pi(2n_\al+\rho_{X\al})\cos^2
\frac{\pi}2(\rho_{X\al})}{\sin\pi(\rho_{X\al})\cos^2\frac{\pi}2(2n_\al+\rho_{X\al})}
%\times
%$$
%$$
\frac{c_1(2n_\al+\rho_{X\al},l_\al)}{c_1(\rho_{X\al},l_\al)}
\frac{c_1(2n_\al+\rho_{X\al},r_\al)}{c_1(\rho_{X\al},r_\al)}\,.
$$
Thus we obtain
\beq{nm}
d_\al(n_\al)=\frac{2n_\al+(\rho_X)_\al}{(\rho_X)_\al}
\frac{c_1(n_\al,l_\al)}{c_1(0,l_\al)}\frac{c_1(n_\al,r_\al)}{c_1(0,r_\al)}\,.
\eq
It is a polynomial of degree $m_\al+m_{2\al}$ (see (\ref{m2}), (\ref{rj})).

$\bullet$ $\bullet$ \emph{The case B}
%$$
%\frac{\G(-\rho_{X\al})\G(\rho_{X\al})}{\G(-(\la,\rho_{X_\al})\G((\la,\rho_{X_\al})}\stackrel{(\ref{gs})}=
%%
%%\frac{\G(-\rho_{X\al})\G(\rho_{X\al})}{
%%\G(-n_\al-\rho_{X\al})\G(n_\al+\rho_{X\al})}\stackrel{(\ref{gs})}=
%%$$
%%$$
%\frac{(2n_\al+\rho_{X\al})\sin\pi(n_\al+\rho_{X\al})}{(\rho_{X\al})\sin\pi(\rho_{X\al})}\,.
%%\frac{n_\al+\rho_{X\al}}{\rho_{X\al}}\,.
%$$

$$
\G(\frac{m_\al}4+\oh+\oh(2n_\al+ \rho_{X\al}))\G(\frac{m_\al}4+\oh -\oh(2n_\al+ \rho_{X\al}))
\stackrel{(\ref{gs1})}=
$$
$$
\frac{\pi (n_\al+ \oh\rho_{X\al})}{\sin\pi (n_\al+ \oh\rho_{X\al})}c_0(n_\al+ \oh\rho_{X\al},l_\al+1)\,,
$$

$$
\G(\frac{m_\al}4+\frac{m_{2\al}}2 -\oh(2n_\al+\rho_{X\al}))\G(\frac{m_\al}4+\frac{m_{2\al}}2 +\oh(2n_\al+\rho_{X\al}))=
$$
$$
\G(r_\al+1-\oh(n_\al+\rho_{X\al}))\G(r_\al+1 +(n_\al+\oh\rho_{X\al}))\stackrel{(\ref{gs7})}=
\frac{\pi(n_\al+\oh\rho_{X\al})}{\sin\pi(n_\al+\oh\rho_{X\al})}c_0(n_\al+\oh\rho_{X\al},r_\al+1)\,.
$$

Then (\ref{mr}) is equal
$$
d_\al(n_\al)=
\frac{2n_\al+\rho_{X\al}}{\rho_{X\al}}
\left(\frac{(2n_\al+\rho_{X\al})\sin\pi(\rho_{X\al})}{(\rho_{X\al})\sin\pi(2n_\al+\rho_{X\al})}\right)^2\times
$$
$$
\frac{c_0(n_\al+ \oh\rho_{X\al},l_\al+1)}{c_0(\oh\rho_{X\al},l_\al+1)}
\frac{c_0(n_\al+\oh\rho_{X\al},r_\al+1)}{c_0(\oh\rho_{X\al},r_\al+1)}
$$
%Using the polynomial $c_0$ (\ref{c0})
 We come finally to the
expression
\beq{3d}
d_\al(n_\al)=\left(\frac{2n_\al+\rho_{X\al}}{\rho_{X\al}}\right)^3
\frac{c_0(n_\al+\oh\rho_{X\al},l_\al)}{c_0(\oh\rho_{X\al},l_\al)}
\frac{c_0(n_\al+\oh\rho_{X\al},r_\al)}{c_0(\oh\rho_{X\al},r_\al)}\,.
%\prod_{k=1}^{l}\frac{k^2-(n+r+1)^2}{k^2-(r+1)^2}
%\prod_{k=1}^{r}\frac{k^2-(n+r+1)^2}{k^2-(r+1)^2}
\eq
%$$
%\frac{(n_\al+1)\ldots(n_\al+l_\al)^2(n_\al+l_\al+1)\ldots(n_\al+2l_\al-1)}{l_\al(2l_\al-1)!}
%\times
%$$
%$$
%\times\frac{(n_\al+1)\ldots(n_\al+r_\al)^2(n_\al+r_\al+1)^4\ldots(n_\al+2r_\al-1)}
%{r_\al(r_\al+1)^3(2r_\al-1)!}
%\,.
%$$
Again,  it is a polynomial of degree $m_\al+m_{2\al}$.
%% (see (\ref{m2}), (\ref{rj})).
%

%%%%%%%%%%%%%%%%%%%%%%%%%%%%%%%%%%%%%%%%%%%%%%%%%%%%%%%%%%%%%%%%%%%%%%%

%%%%%%%%%%%%%%%%%%%%%%%%%%%%%%%%%%%%%%%%%%%%%%%%%%%%%%%%%%%%%%%%%%%%%%

\section{The type I zeta functions for symmetric spaces rank one}

\setcounter{equation}{0}

Here we write down the explicit expressions for the
type I zeta functions for the rank one symmetric spaces and discuss their relations to  the Riemann zeta function.
Riemann symmetric spaces of rank one are classified by the multiplicities $m_\al$ and $m_{2\al}$.
Their dimensions are
\beq{1}
\dim\,X=1+m_\al+m_{2\al}\,.
\eq

It follows from Tables 1. 2. that there are three infinite classical series
 (the real, complex and quaternionic projective spaces) and one exceptional case
 (the Cayley Projective Plane FII)
\begin{center}
\begin{tabular}{|c|c|c|c|c|l|}
  \hline
  \hline
 $\mR$& BDI* &
$S^m=SO(m+1)/SO(m)$ &$ m_\be=(m-1)$& $m_{2\be}=0 $&$\rho=(m-1)/2$
\\
\hline
$\mC$&   AIII & $\mC P^m=
 SU(m+1)/S(U(m)\times U(1))$  &$ m_\be=2(m-1)$
 & $m_{2\be}=1 $ &$\rho=m$
\\
\hline
$ {\mathbb H} $&CII & ${\mathbb H}P^m=Sp(m+1)/Sp(m)\times Sp(1)$ &$ m_\be=4(m-1)$& $m_{2\be}=3 $
&$\rho=2m+1$
\\
\hline
  & FII & $F_4/SO(9)$ & $m_\be=8$ & $m_{2\be}=7$&$\rho=11$
  \\
\hline
\hline
\end{tabular}

\bigskip
\textbf{Table 3. Rank one type I symmetric spaces}

\end{center}
The type I zeta functions (\ref{wi1}) takes the form
\beq{wi4}
\zeta^I_{U/K}(s)=\sum_{n\in\mZ^+}\frac{1}{(d_{U/K}(n))^s}\,.
\eq

%%%%%%%%%%%%%%%%%%%%%%%%%%%%%%%%%%%%%%%%%%%%%%%%%%%%%%%%%%%%%%%%%%%%%

\subsection{Calculations of dimensions}
Let's consider each case separately. We use expressions for $d(n)$ obtained in
previous Section.

%%%%%%%%%%%%%%%%%%%%%%%%%%%%%%%%%%%%%%%%%%%%%%%%%%%%%%%%%%%%%%%%%%%%%%%%%%%%
\subsubsection{Real case $S^{m}=SO(m+1)/SO(m)$, ($m_{2\al}=0$)}
%$$
%G=SO(m,1)\,,~~U=SO(m+1)\,,~~K=SO(m)\,,
%%~~ Lie(G)=\gg_0+m_\al\gg_\al\,,~~
%%\gg_0=Lie(SO(m-1))\,,
%$$
$$
m_\al=m-1\,,~l=(m-1)/2\,, `m_{2\al}=0\,,~~\rho_{S^{m}}=(m-1)/2\,,
%~~\imath((\la+\rho_{S^{m}}),\al_0)=n+\frac{m-1}2\,.
 $$

For  $m$ odd
%\beq{m1}
$(m_\al=2l\,,~~l\geq 1\,,~~ \rho_X=l)$
%\eq
from (\ref{di4}) we find
\beq{r14}
d_{S^m}(n)=\frac{2n+m-1}{m-1}\cdot\frac{(n+1)\ldots(n+m-2)}{(m-2)!}
%\prod_{k=0}^{l-1}\frac{k^2-(n+l)^2}{k^2-l^2}=\frac{c_0(n,l)}{c_0(0,l)}
\,.
\eq

For $m$ even ($m_\al=2l+1$, $l\geq 1$) we use (\ref{di5})
\beq{r13}
d_{S^m}(n)=
\frac{(n+1)\ldots(n+\frac{m}2-1)(n+\frac{m}2-\oh)(n+\frac{m}2))\ldots(n+m-2)}{\frac{m-1}2(m-2)!}\,.
\eq
%\frac{2n+2l+1}{2l+1}
%\prod_{k=1}^{l}\frac{(k-\oh)^2-(n+l+1/2)^2}{(k-\oh)^2-(l+1/2)^2}=
%\frac{2n+2l+1}{2l+1}\cdot\frac{c_1(n,l)}{c_1(0,l)}
%\,.
%\eq

In particular,
$$
d_{S^2} (n)=2n+1\,,
$$
$$
d_{S^3} (n)=
(n+1)^2\,.
$$
$$
d_{S^4} (n)=\frac{(n+1)(n+1+\oh)(n+2)}{ 3}\,.
$$
$$
d_{S^5} (n)=
\frac{(n+1)(n+2)^2(n+3)}{2^23}\,.
$$
$$
d_{S^6} (n)=\frac{(n+1)(n+2)(n+2+\oh)(n+3)(n+4)}{ 5\cdot 4\cdot 3}\,.
$$
$$
d_{S^7} (n)=\frac{(n+1)(n+2)(n+3)^2(n+4)(n+5)}{2^33^25}\,.
$$

%%%%%%%%%%%%%%%%%%%%%%%%%%%%%%%%%%%%%%%%%%%%%%%%%%%%%%%%%%%%%%%%%%%%%%%%%%%%
\subsubsection{Complex case. $\mC P^{m}=SU(m+1)/S(U(m)\times U(1))$}

The root system in this case is
 $$
 \begin{array}{cccc}
 m=1 & C_1 & m_\al=1 &  m_{2\al}=0\\
   m>1 & BC_1 & m_\al=2(m-1) & m_{2\al}=1\,,~~j=0\,.
 \end{array}
$$
In the first case $\mC P^1\sim S^2$ and $d_{\mC P^1 }(n)=2n+1$.

The case A (\ref{rj}) corresponds to the odd $m$:
$$
l=\frac{m-1}2\,,~~m\geq 3\,,~~r=l\,.
$$
Then from (\ref{nm}) we have
$$
d(n)=\frac{(n+1)^2\ldots(n+m-1)^2(n+\frac{m}2)}{((m-1)!)^2\frac{m}2}\,.
$$

The case B (\ref{rj}) corresponds to the even $m$:
$$
l=\frac{m-2}2\,,~~\,,m\geq 4\,,~~r=l\,, \,.
$$
$$
d(n)=
\frac{(n+1)\ldots(n+\frac{m-2}2-1)(n+\frac{m-2}2)^2(n+\frac{m}2)\ldots(n+m-3)}{\frac{m-2}2(m-3)!}
\times
$$
$$
\times\frac{(n+1)\ldots(n+\frac{m-2}2)^2(n+\frac{m}2)^4\ldots(n+m-3)}
{\frac{m-2}2(\frac{m-2}2+1)^3(m-3)!}
\,.
$$

In particular,
$$
d_{\mC P^2}(n)=(n+1)^3\,,
$$
$$
d_{\mC P^3}(n)=\frac{(n+1)^2(n+1+\oh)(n+2)^2}{2^23}\,,
$$
$$
d_{\mC P^4}(n)=\frac{(n+2)^3(n+1)^2(n+3)^2}{2^33^2}\,.
$$

%%%%%%%%%%%%%%%%%%%%%%%%%%%%%%%%%%%%%%%%%%%%%%%%%%%%%%%%%%%%%%%%%%%%%%%%%%%%
\subsubsection{Quaternionic case. $\mathbb{H}P^{m}=Sp(m+1)/Sp(m)\times Sp(1)$}

$$
 BC_1\,,~~ m_\al=4(m-1)\,,~~ m_{2\al}=3\,,~~j=1\,.
 $$
It is the case A (\ref{rj})
$$
l=m-1\,,~~ r=l+j=m\,.
$$
From (\ref{nm}) we have
$$
d(n)=
\frac{(n+1)\ldots(n+2m-2)}{(2m-2)!}\times
\frac{(n+1)\ldots(n+m)(n+m+\oh)(n+m+1)\dots((n+2m)}{(m+\oh)(2m)!}
%\frac{2n+2m+1}{2m+1}\frac{c_1(n,m-1)}{c_1(0,m-1)}\frac{c_1(n,m)}{c_1(0,m)}\,.
$$
In particular,
$$
d_{\mathbb{H}P^{1}}(n)=\frac{(n+1)(n+1+\oh)(n+2)}{ 3}\,,
$$
$$
d_{\mathbb{H}P^{2}}(n)=\frac{(n+1)^2(n+2)^2(n+2+\oh)(n+3)(n+4)}{2^3\cdot 3\cdot5}\,,
$$
%%%%%%%%%%%%%%%%%%%%%%%%%%%%%%%%%%%%%%%%%%%%%%%%%%%%%%%%%%%%%%%%%%%%%%%%%%%%
\subsubsection{Cayley Projective Plane
 $FII=F_4/SO(9)$}

This case corresponds the variant A (\ref{rj}) with $l=2$ and $j=3$.
Then putting in (\ref{rj}) $r=l+j=5$ we obtain
$$
d_{FII}(n)=\frac{2n+11}{11}\left(\frac{c_1(n,2)}{c_1(0,2)}\right)^2\frac{c_1(n,5)}{c_1(0,5)}=
$$
$$
\frac{2n+11}{11}\times\frac{(n+1)(n+2)(n+3)(n+4)}{4!}\times
\frac{(n+1)(n+2)\ldots(n+10)}{10!}=
$$
$$
\frac{(n+1)^2(n+2)^2(n+3)^2(n+4)^2(n+5)(n+5+\oh)(n+6)\ldots(n+10)}
{2^2\cdot 3\cdot 11!}
$$

Summarising, we write the general formula for dimension of rank one symmetric spaces
\beq{di}
d_X(n)=
%D_X(n+\rho_X/2)
c\prod_{k=1}^M(n+\ka_k)^{\xi_k}\,,
\eq
where
\beq{xi}
%\pi_0=1\,,{\rm or\,}\,3\,,~~
\xi_k=1\,,\,2\,,\,3\,;~~\sum_{k=1}^M\xi_k=\dim\,X-2\,,
\eq
\beq{ka1}
a)\,\ka_k=k\,,~{\rm or~}b)\,\ka_k= k-1+\oh\,,
\eq
and $c$ is an $n$ independent constant.
%$D_X$ is the normalization factor: $D_X=d^{-1}_X(0)$.

 For example, for FII
$$
M=10\,,~c =\frac{1}{ 2^2\cdot 3\cdot 11!}\,,~~
\xi_1=\ldots=\xi_4=2\,,~~\xi_5=\ldots=\xi_{10}=1\,,~\ka_6=5+\oh\,.
$$

%%%%%%%%%%%%%%%%%%%%%%%%%%%%%%%%%%%%%%%%%%%%%%%%%%%%%%%%%%%%%%%%%%%%%%%%%%%%%%%%
%%%%%%%%%%%%%%%%%%%%%%%%%%%%%%%%%%%%%%%%%%%%%%%%%%%%%%%%%%%%%%%%%%%%%%%%%%%%%%%%%%%
\subsection{Values of Type I rank one zeta functions at integer arguments }

It follows from (\ref{di}) that the type one zeta functions for the rank one spaces takes the form
\beq{wi5}
\zeta^I_{X}(s)=c^{-s}\sum_{n\in\mZ^+}\prod_{k=1}^M(n+\ka_k)^{-s\xi_k}\,.
%(n+\rho_X/2)^{-s}\prod_{k=1}^M(n+\k)^{-s\xi_k}\,.
\eq
%$$
% D_X^{-s}\sum_{n\in\mZ^+}\prod_{k=1}^M(n+\ka_k)^{-s\xi_k}   \,,
%$$
%$$
%(\ka_0=\rho_X/2\,,~\xi_0=1)\,,~~(\ka_k=1,2,\ldots\,,~ \xi_k=1\,,2\,,3~for~k>0),.
%$$
We are interesting in values of $\zeta^I_{X}(s)$ for integer values of $s$.
In particular for $s=2g-1$ we come to the partition functions of the YM theories
in the limit $\varrho\to 0$
over the Riemann surface $\Si_g$
with the gauge groups mentioned in Table 3.

We consider a generating function for values of $\zeta^I_{X}(s)$ ($s\in\Bbb N$).
%
%In this case
%denote $s\xi_k=1+\pi_k$, $\pi_k\in\mN$.
Denote
$$
\Pi=\left\{
\begin{array}{ccc}
  \pi_1   &\cdots & \pi_M\\
     \kappa_1   &\cdots & \kappa_M
\end{array}\right\}\,,~\pi_j,\in \Bbb N\,,~\kappa_j\in \Bbb N\,,~or~\Bbb N+\frac{1}2,.
%\kappa_j
$$

For $M>1$ we shall study the series
\beq{pi}
\zeta[\Pi]=\sum_{n=1}^\infty \prod_{j=1}^M (n+\kappa_j)^{-(1+\pi_j)}\,.
\eq
Our case corresponds to $\pi_k=s\xi_k-1$ and $\xi_k$ and $\ka_k$ are defined in (\ref{xi})
and (\ref{ka1}).

Pass  to the generating series
$$
\clZ(\begin{array}{ccc}
     \kappa_1   &\cdots & \kappa_N
\end{array};\begin{array}{ccc}
     T_1   &\cdots & T_M
\end{array})
=\sum_{\begin{array}{ccc}
     \pi_1   &\cdots & \pi_M
\end{array}} \zeta[\Pi]
  T_1^{\pi_1}   \cdots T_M^{ \pi_M}=
$$
$$
\sum_{n=1}^\infty \prod_{j=1}^M (n+\kappa_j-T_j)^{-1}
$$

 For $M=1$, $\ka\in\mN$ (\ref{ka1}) and $\pi=p-1$ (\ref{pi}) becomes
%$p\neq 1$
$$
%\clZ\left[\begin{array}{c}p-1\\ \kappa\
%\end{array}\right]=
\zeta\left[\begin{array}{c}p-1\\ \kappa\
\end{array}\right]=\zeta(p)-\sum_{n=1}^\kappa n^{-p}\,,
$$
where $\zeta(p)$ is the Riemann zeta function.
 For $p=1$ the series
diverges but  the difference
$$
\zeta\left[\begin{array}{c}0\\
 \kappa\
\end{array}\right]-\zeta\left[\begin{array}{c}0\\ \lambda\
\end{array}\right]
$$
is correctly  defined and is equal to $ \sum_{n=\lambda +1}^\kappa n^{-1}$. Hence the difference $\clZ(\kappa;T)-\clZ(\lambda;S)$ is also well-defined.

For $\ka=k+\oh$ (\ref{ka1}) (\ref{pi}) becomes
$$
\zeta\left[\begin{array}{c}p-1\\ \kappa\
\end{array}\right]=\zeta(p,\oh)-\sum_{n=1}^\kappa n^{-p}\,,
$$
where
$$
\zeta(p,\oh)=\sum_{n=1}^\infty\f1{(n+\oh)^p}=(2^p-1)\zeta(p)\,.
$$

We need the equality
$$
\prod_{n=1}^M (a_n)^{-1}=\sum^M_{\nu=1}\left(
\prod_{\mu\neq \nu} (a_\mu-a_\nu)^{-1}\right)(a_\nu)^{-1}
$$
It can be checked by considering the limit $\lim (a_\mu- a_\nu)\to 0$.
Our main formula follows from this equality
\beq{mf}
Z(\begin{array}{ccc}

     \kappa_1   &\cdots & \kappa_N
\end{array};\begin{array}{ccc}

     T_1   &\cdots & T_N
\end{array})=
\eq
$$
\sum_{\nu}
\left(
\prod_{\mu\neq \nu} ((\kappa_\mu-\kappa_\nu)-(T_\mu-T_\nu))^{-1}\right)Z( \kappa_\nu,T_\nu)\,,
%\begin{array}{c}
%
%     \kappa_\nu
%\end{array};\begin{array}{ccc}
%
%     T_1   &\cdots & T_N
%\end{array})
$$
where
$$
Z( \kappa_\nu,T_\nu)=\sum_{n=1}^\infty\f1{(n+\ka_\nu+T_\nu)^{1+\pi_\nu}}\,.
$$

 The right-hand side of the equality is well-defined as
 $$\sum_{\nu}
\left(
\prod_{\mu\neq \nu} ((\kappa_\mu-\kappa_\nu)-(T_\mu-T_\nu))^{-1}\right)=0$$
%\end{document}
%$$\frac{1}{(-n)^a}\frac{1}{(n+1)^b}=\sum_{h=0}^{a-1}
%\frac{(b+h-1)!}{h!(b-1)!}\frac{1}{(-n)^{a-h}}+\sum_{k=0}^{b-1}
%\frac{(a+k-1)!}{k!(a-1)!}\frac{1}{(n+1)^{b-k}}$$
%
%$$\frac{1}{(-2l)^{2r}}\frac{1}{(2l+1)^{2r}}\frac{1}{(2l+2)^{2r}}=$$
%
%$$ =\sum_{h=0}^{2r-1}
%\frac{(2r+h-1)!}{h!(2r-1)!}\frac{1}{(-2l)^{2r-h}}\frac{1}{(2l+2)^{2r}}+\sum_{k=0}^{24-1}
%\frac{(2r+k-1)!}{k!(2r-1)!}\frac{1}{(2l+1)^{2r-k}}\frac{1}{(2l+2)^{2r}}=$$
%$$\sum_{h=0}^{2r-1}
%\frac{(2r+h-1)!}{h!(2r-1)!}
%\left(
%\sum_{i=0}^{2r-h-1}
%\frac{(2r+i-1)!}{i!(2r-1)!}\frac{1}{(-2l)^{2r-h-i}2^{2r+i}}+\sum_{j=0}^{2r-1}
%\frac{(2r-h+j-1)!}{j!(2r-h-1)!}\frac{1}{(2l+2)^{2r-j}2^{2r-h+j}}
%\right)+
%$$
%
%$$+\sum_{k=0}^{2r-1}(-1)^{2r-k}
%\frac{(2r+k-1)!}{k!(2r-1)!}
%\left(
%\sum_{i=0}^{2r-k-1}
%\frac{(2r+i-1)!}{i!(2r-1)!}\frac{1}{(-2l-1)^{2r-k-i}}+
%\sum_{j=0}^{2r-1}
%\frac{(2r-k+j-1)!}{j!(2r-k-1)!}\frac{1}{(2l+2)^{2r-j}}
%\right)
%$$
%
%
%\frac{1}{(-2l)^{2r-h}}\frac{1}{(2l+2)^{2r}}
%
%
%
%

%%%%%%%%%%%%%%%%%%%%%%%%%%%%%%%%%%%%%%%%%%%%%%%%%%%%%%%%%%%%%%%

%%%%%%%%%%%%%%%%%%%%%%%%%%%%%%%%%%%%%%%%%%%%%%%%%%%%%%%%%%%%%%
\begin{small}
\vspace{2mm}
{\bf Acknowledgments.} We are grateful to A.Gorsky and A.Mironov
 for useful discussions.
The work of M.O. was carried out in part within the state assignment of
NRC Kurchatov Institute.
The work of A.L. was supported by the RFBR  grant 21-41-09011.

\end{small}

\end{document}